\documentclass[utf8]{FrontiersinHarvard_arXiv} %

\usepackage{url,hyperref,lineno,microtype,subcaption}
\usepackage[onehalfspacing]{setspace}
\usepackage{amsmath, amsthm, amssymb, amsfonts}

\def\keyFont{\fontsize{8}{11}\helveticabold }
\def\firstAuthorLast{Lochner {et~al.}} %
\def\Authors{Stephan Lochner\,$^{1,*}$, Daniel Honerkamp\,$^{2}$, Abhinav Valada\,$^{2}$ and Andrew D. Straw\,$^{1,3}$}
\def\Address{$^{1}$Institute of Biology I, University of Freiburg, Germany \\
$^{2}$Department of Computer Science, University of Freiburg, Germany \\
$^{3}$Bernstein Center Freiburg, University of Freiburg, Germany}
\def\corrAuthor{Stephan Lochner}

\def\corrEmail{stephan.lochner@biologie.uni-freiburg.de}

\begin{document}
\onecolumn
\firstpage{1}
\title[Reinforcement Learning for Insect Navigation]{Reinforcement Learning as a Robotics-Inspired Framework for Insect Navigation: From Spatial Representations to Neural Implementation.}

\author[\firstAuthorLast ]{\Authors} %
\address{} %
\correspondance{} %

\extraAuth{}%

\maketitle

\begin{abstract}
    \section{}
    Bees are among the master navigators of the insect world.
    Despite impressive advances in robot navigation research, the performance of these insects is still unrivaled by any artificial system in terms of training efficiency and generalization capabilities, particularly considering the limited computational capacity.
    On the other hand, computational principles underlying these extraordinary feats are still only partially understood.
    The theoretical framework of reinforcement learning (RL) provides an ideal focal point to bring the two fields together for mutual benefit.
    In particular, we analyze and compare representations of space in robot and insect navigation models through the lens of RL, as the efficiency of insect navigation is likely rooted in an efficient and robust internal representation, linking retinotopic (egocentric) visual input with the geometry of the environment.
    While RL has long been at the core of robot navigation research, current computational theories of insect navigation are not commonly formulated within this framework, but largely as an associative learning process implemented in the insect brain, especially in the mushroom body (MB).
    Here we propose specific hypothetical components of the MB circuit that would enable the implementation of a certain class of relatively simple RL algorithms, capable of integrating distinct components of a navigation task, reminiscent of hierarchical RL models used in robot navigation.
    We discuss how current models of insect and robot navigation are exploring representations beyond classical, complete map-like representations, with spatial information being embedded in the respective latent representations to varying degrees.

    \tiny
    \keyFont{ \section{Keywords:} insect navigation, reinforcement learning, robot navigation, mushroom bodies, spatial representation, cognitive map, internal world model} %
\end{abstract}

\section{Introduction}
\label{sec:intro}

The purpose of this paper is two-fold: First, we offer a perspective that links our current understanding of spatial navigation by insect navigation researchers together with that of robotics researchers.
We do this largely with the help of the theoretical framework of reinforcement learning (RL), which is a central theme in modern robotics research but has so far had relatively little impact in the field of insect navigation or more broadly in insect learning.
We focus on an analysis of spatial representations in current robot and insect navigation models through the lens of RL.
Second, we propose neural mechanisms by which the anatomy and physiology of the insect brain may implement RL-like learning, and finally offer a hypothesis on how recent models describing distinct components of insect navigation can be combined into a hierarchical RL model.

Reliably navigating the world in order to acquire essential resources while avoiding potentially catastrophic threats is an existential skill for many animals.
In the insect world, central-place foragers like many bee and ant species stake the survival of the entire colony on individuals' ability to return to the nest after extensive foraging trips.
Their remarkable navigational capabilities allow them to do so after only a few learning flights or walks under vastly varying environmental conditions.
This is so far unrivaled by any artificial autonomous system.
Recent years have seen substantial advances in understanding the underlying mechanisms of insect navigation (INav), tentatively converging on what was coined the `insect navigation base model' (INBM) in a comprehensive review by \citet{webbInternalMapsInsects2019}.

This model and its components \textendash{} rooted in a rich history of behavioral experiments, modeling, and the neuroanatomy of the insect brain \textendash{} possess substantial explanatory power and offer a mechanistic, bottom-up picture of navigation.
Nevertheless, it is only implicitly related to the high-level objective of efficiently exploiting the resources provided by the environment.
On the other hand, robot navigation (RNav) research is driven by the practical goal of enabling robots to perform specific spatial tasks, making reinforcement learning a dominant theoretical framework:
An RL agent is trained to optimize its interaction with the environment by accumulating positive rewards while avoiding punishment (negative rewards), which it achieves by learning a specific policy: what is the optimal action to take, given the agent's current state?
In contrast to other training paradigms, RL requires no additional external supervision.
If the task involves a spatial component, this implies learning a navigational strategy which is optimal for achieving the high-level task.

Successful and reliable navigation depends on a robust and efficient choice of the agent's internal representation of its environment (\emph{mapping}) and its own relative pose therein (\emph{localization}), which will be derived from sensory input but otherwise arbitrarily complex.
This \emph{spatial representation} of sensory input then serves as the basis to determine a sequence of suitable actions to accomplish a certain objective (\emph{planning}).
Until recently, the dominant approach in RNav decoupled the question of finding a suitable spatial representation from the planning phase in modular architectures: a fixed, feed-forward architecture (usually some variety of `simultaneous localization and mapping', SLAM, see \citealt{fuentes-pachecoVisualSimultaneousLocalization2015} for a review) is used to infer an \emph{explicit} spatial representation from sensory input based on which a policy is optimizing its actions using e.g. classical planners or learned RL methods.
In current research, however, end-to-end learning approaches are increasingly gaining traction, where differentiable neural network architectures are trained to learn policies directly from the sensory input.
In order to do so efficiently, these networks usually form \emph{latent} spatial representations within hidden layers of the network as an intermediate step.
These \emph{latent representations} are not pre-determined but learned in order to most efficiently solve the navigation task within the constraints of a specific network architecture\footnote{Although other training paradigms with varying degrees of supervision are being used as well, most of the models discussed can be trained in an end-to-end RL fashion, and we will interpret their latent representations within a common RL framework for conceptual clarity.
}.

Spatial representations can be further characterized by their `geometric content', i.e. how much of the geometric structure of the environment is encoded in the spatial representation:
as a biological example, spatial firing fields of hippocampal place-cells \citep{okeefeHippocampusSpatialMap1971} in mammals tile the entire (accessible) environment of the agent, giving rise to a dense spatial representation akin to grid-like spatial maps used in RNav, although geometry is generally not thought to be preserved accurately.
On the other hand, similarity gradients on a retinotopic (pixel-by-pixel) level, as proposed for visual navigation in insects \citep{zeilCatchmentAreasPanoramic2003}, carry no geometric information about the environment at all.
This mirrors the long-standing debate among insect navigation researchers whether insects use cognitive maps for navigation \citep{dheinCognitiveMapDebate2023}.
Following a common negative characterization of an animal \emph{without} a cognitive map:
\emph{"At any one time, the animal knows where to go rather than where it is [\dots]"} \citep{hoinvilleOptimalMultiguidanceIntegration2018},
we can restate the question in the language of RL as follows:
\begin{quote}
    \emph{What is the geometric content (`where the animal is') of the - latent or explicit - spatial representation (`what the animal knows') of the RL agent?}
\end{quote}

Free of anatomical and physiological constraints, recent RNav research has produced a plethora of end-to-end learned navigation models with different architectures and policy optimization routines.
The resulting pool of `experimentally validated' latent spatial representations can serve as theoretical guidance when thinking about the way space is represented in the insect brain for successful navigation \textendash{} both in terms of behavioral modeling and in the experimental search for neural correlates of such representations.
Conversely, evidence about certain components of spatial representations in insects, like the existence of spatial vectors encoded in the brain, may guide the design of network architectures for artificial agents.
To this end, we will analyze what geometric information is represented (Sec.~\ref{sec:latentspaces}) and how it is represented in recently successful robot navigation models (Sec.~\ref{sec:spatial_rep_robo}) and in the `insect navigation base model' (Sec.~\ref{sec:spatial_rep_insect}).

Going beyond the conceptual considerations outlined above, the question naturally arises whether a link between insect navigation and RL can be established on a more fundamental level.
After a brief formal introduction to RL (Sec.~\ref{subsec:RL_formalim}),
we will investigate how the neuroanatomical components involved in the insect navigation base model, the mushroom bodies (MB) and central complex (CX), could support computations similar to certain simple RL algorithms like SARSA or Q-learning.
We present current models of MB neural computation (Secs.~\ref{subsubsec:MB_canonical} and \ref{subsubsec:MB_RPE}), to propose specific hypothetical neural connections and their physiological properties by which the models could be augmented to support temporal difference learning.
In
(Sec.~\ref{sec:RL_insect_navigation}), we sketch a how such a model could be integrated with a recent MB based visual homing model \citep{wystrachNeuronsPremotorAreas2023} into a full RL-based visual navigation model.
Finally, we discuss in Sec.~\ref{sec:discussion}, what kind of spatial representation would result from such a model, its implication for the cognitive map debate and how it aligns with models currently used in robot navigation.

\section{Representations of Space from a RL Perspective}
\label{sec:latentspaces}

In robot navigation, the problem of navigation has traditionally been partitioned into the subtasks of \emph{localization, mapping} and \emph{planning}.
Mapping and localization operations take (potentially multimodal) sensory input to infer a map of the environment and the agent's pose.
It has long been acknowledged that the localization problem is most easily solved by reference to locations of salient landmarks in the world - i.e. a map - and conversely, constructing a coherent map requires accurate estimates of the agent's pose.
This led to the breakthrough of a suite of techniques collectively known as \emph{`simultaneous localization and mapping'} (SLAM) \citep{mur-artalORBSLAM2OpenSourceSLAM2017, engelLSDSLAMLargeScaleDirect2014, endres2012evaluation, fuentes-pachecoVisualSimultaneousLocalization2015}.
Most SLAM techniques combine landmark/feature recognition with odometry to maintain a joint (often probabilistic) representation of the environment and the agent's pose therein, which we will refer to as the \emph{spatial representation} $\phi \in \Phi$ of the sensory input $v \in \mathcal{V}$.
We denote sensory input with $v$, since the paper will focus on visual navigation, for simplicity.
Multimodal input spaces are of course possible and highly relevant for a realistic understanding of insect navigation\footnote{Note the definition of `visual input' may differ between INav and RNav and go beyond pure RGB pixel values.
    For example, insects can visually infer compass cues from celestial polarization patterns, while depth information obtained from RGB-D cameras is often used RNav}.
Based on $\phi$, the planning stage then determines a sequence of actions $a \in \mathcal{A}$ in order to achieve the objective of the navigation task (see Fig.~\ref{fig:NavigationMDP}A).
This in turn can be based on planning-based or learned methods.
In the following, we analyze the spatial representations $\Phi$ found in current robot and insect navigation models.
Since these representations differ in many aspects, we first define two dimensions along which our analysis is structured: \emph{`What is represented?
    '} and \emph{`How is it represented?'}.

\begin{figure*}
    \centering
    \includegraphics[width=1.0\textwidth]{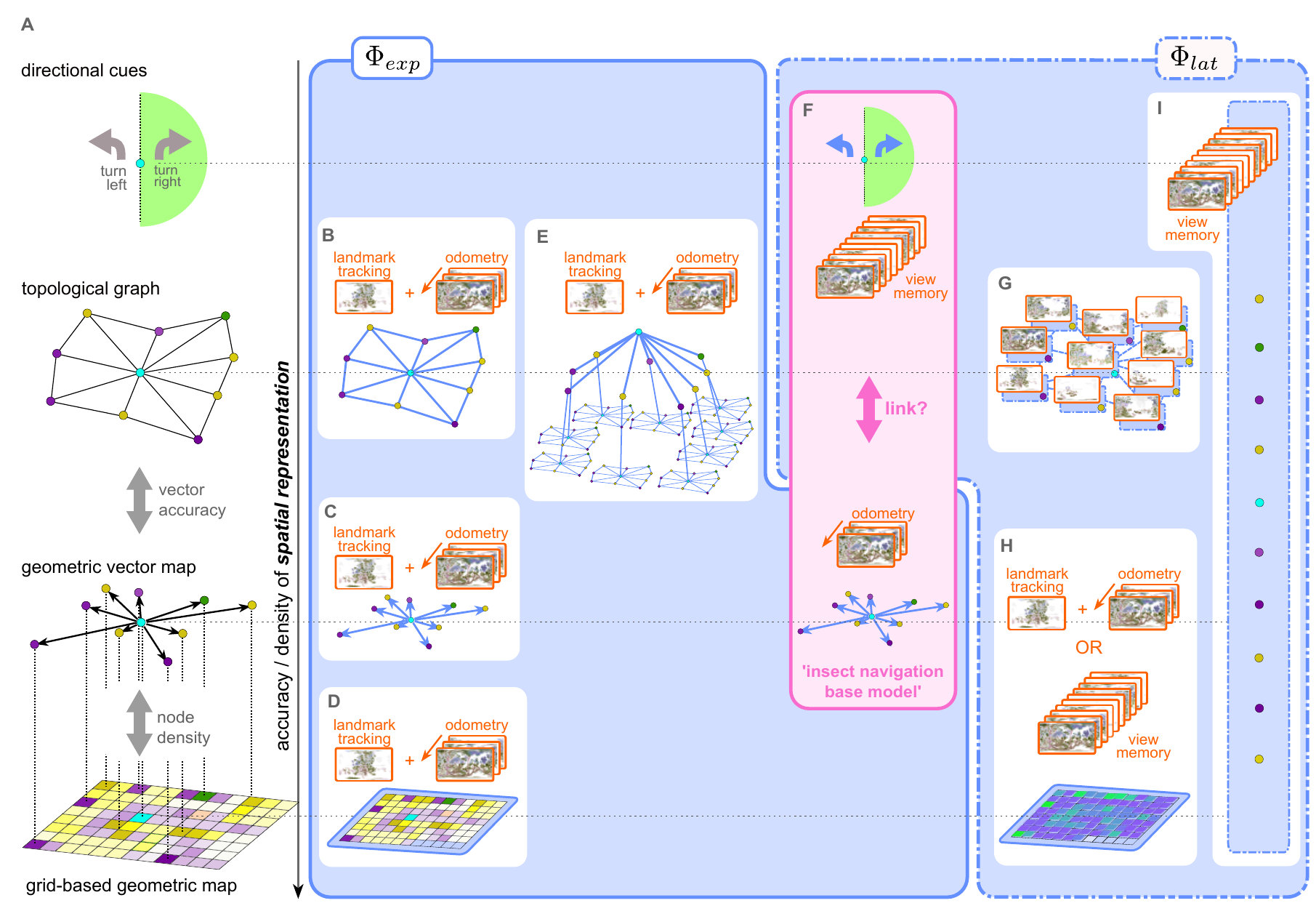}
    \caption{(A) representations of space with varying degrees of geometric accuracy (geometric content) (B-H) explicit and latent spatial representations in robot and insect navigation. (B-D) visual SLAM methods use a combination of odometry and landmark/feature tracking and matching to construct explicit topological maps (B), vector/landmark (C), or grid-based (D) maps.
        (E) Scene-graphs \citep{hughes2022hydra,werbyHierarchicalOpenVocabulary3D2024,honerkampLanguageGroundedDynamicScene2024} construct hierarchical, object-centric graph representations to disassemble large scenes into regions, objects, etc, based on SLAM-like mapping approaches.
        Edges usually encode predefined, i.e. explicit, relational, and semantic attributes.
        (F) the insect navigation base model uses two distinct mechanisms which result in two spatial representations: an explicit vector map, built from visual odometry alone without the need for landmark recognition and mapping, and latent directional cues learned from a view memory.
        It is not clear if and how these two are linked to form a unified representation of space.
        (G-I) more recent approaches in robot navigation use latent representations.
        (G) topological latent representations \citep[e.g.
            RECON]{shahRapidExplorationOpenWorld2023}
        (H) grid-based latent representations can be built from an explicit view-memory based architecture using RNNs \citep[e.g. MapNet:]{henriquesMapNetAllocentricSpatial2018} or a learnable mapping module as in CMP \citep{guptaCognitiveMappingPlanning2019}.
        While the grid structure is fixed, the content of the grid (i.e. the interpretation of the values stored in the grid) is latent, as opposed to an explicit occupancy or semantic grid map (D).
        (I) Unstructured memory based approaches like SMT~\citep{fangSceneMemoryTransformer2019} or unstructured RNNs \citep[like][]{beechingDeepReinforcementLearning2019} learn a completely abstract latent spatial representation, which defies classification along the vertical axis in the figure.
    }
    \label{fig:space_latent_rep}
\end{figure*}

\subsection{What is Represented?
    The Geometric Content of the Spatial Representation $\phi$
}
\label{subsec:geom_content}
Confining our discussion to 2D, the simplest, `geometrically perfect' representation of the environment could be imagined as an infinitely extended and infinitesimally spaced grid, filled with binary `occupancy' values.
While it may prove useful to enrich the map with layers of meaning (object categories, valuations, etc.) by adding semantic channels, the geometric information is captured fully by this single layer\footnote{Depending on the choice of origin and orientation of the grid axes, this map could be either \emph{allocentric} or \emph{egocentric}.
    Any sensory (visual) input is \emph{by definition} egocentric.
}.
Any practical representation of space, however, must be an abstraction of this ideal to varying degrees, trading off density and geometric accuracy for improved coding efficiency and storage capacity (see Fig.~\ref{fig:space_latent_rep}A):
\subsubsection{Grid-based Maps}
The most straightforward simplification is a grid with finite extent and resolution.
Grid-based occupancy maps have a long history in SLAM approaches to robot navigation \citep[e.g.][]{gutmann3DPerceptionEnvironment2008, mur-artalORBSLAM2OpenSourceSLAM2017,engelLSDSLAMLargeScaleDirect2014, endres2012evaluation} where a probabilistic occupancy grid map is predicted from a time series of observations, as a joint estimate for both the agent's position (localization) and layout of the environment (mapping).
More recent methods use hierarchical multi-scale approaches~\citep{zhu2021nice}, neural radiance fields \citep{rosinol2023nerf} or Gaussian splatting \citep{matsuki2023gaussian} for highly accurate reconstructions.
Grid-based maps are both spatially dense and geometrically faithful.
The first aspect of such a representation is also reflected in the spatial firing fields of the so-called place-cells in the mammalian hippocampus \citep{okeefeHippocampusSpatialMap1971}.
For example, \citet{rich2014large} show that place field density is uniform across a given environment, indicating that all regions of the environment are represented.
There is, however, no hard evidence that place-cell representation preserves the geometry of the environment.
Furthermore, recent interpretations \citep{fentonRemappingRevisitedHow2024} of observed `remapping' of spatial fields view place-cell like firing patterns as particular projections of conjoint, collective population activity on a neural manifold, as opposed to the original view, where spatial position was thought to be encoded by dedicated single-neuron activity.
It is currently unknown whether insects also possess neuronal populations with similar place-cell like activity.
We discuss potential candidate cell types in the insect brain in Sec.~\ref{sec:discussion}.

While more classical approaches focus on binary occupancy or probabilistic occupancy encodings as inputs to motion planners, learning based methods have also encompassed higher-dimensional contexts such as semantics~\citep{waniMultiONBenchmarkingSemantic2020, schmalstieg2022learning,younes2023catch} or potential functions~\citep{ramakrishnan2022poni} as additional channels in these maps.

\subsubsection{Vector Maps}
As a next level of abstraction useful for sparsely populated maps, one could store only the grid indices of occupied cells, instead of an occupancy value for every cell.
Increasing the accuracy by replacing grid indices with actual (Cartesian) coordinates with respect to some common origin, we arrive at a vector map, in which geometric relations in the world are represented by relative vectors between salient locations (vector nodes).
\citet{stemmler2015connecting} show how a vector-like spatial representation can be decoded from grid-cell\footnote{Grid-cells have (spatially) periodic firing fields, which can serve as a basis for representing spatial vectors.
    This is not to be confused with a grid-like spatial representation, which would be more compatible with (non-periodic) place-cell like activity, where the receptive field corresponds to a particular position within the spatial grid.
} activity in the mammalian medial enthorinal cortex by combining populations representing different spatial scales.
As we discuss below, the insect navigation base model assumes a vector-based representation of the global geometry of the environment, using a different neural implementation based on phasors \citep{stoneAnatomicallyConstrainedModel2017, lyuBuildingAllocentricTravelling2022}
\subsubsection{Topological Graphs}
If the vector information between connected nodes becomes inaccurate, the vector map gradually loses geometric information and transforms into a topological map, to the extreme case where nodes are connected only by binary `reachability' or `traversability' values.
A less extreme case would be a `weighted graph' representation, where edge weights could represent the Euclidean (or temporal) distance between nodes, preserving some geometric information, but not enough to uniquely reconstruct the map.
Besides the obvious advantage of memory efficiency, a topological representation may be preferred over geometric maps (as argued for by \citealt{warrenWormholesVirtualSpace2017} in humans) for a different reason: It is more robust to inaccurate or corrupted measurements and therefore a more reliable representation of the coarse structure of the environment, which can then be combined with other mechanisms for local goal finding.
Many outdoor navigation approaches in RNav construct topological graphs of the environment \citep[e.g.][]{shahRapidExplorationOpenWorld2023,shahViKiNGVisionBasedKilometerScale2022, engelLSDSLAMLargeScaleDirect2014}.
The gradual transition between the map types described above is illustrated in Fig.~\ref{fig:space_latent_rep}A.

\subsubsection{Scene-Graphs}
As another alternative to dense maps, scene graphs (Fig.~\ref{fig:space_latent_rep}E) have arisen as sparse environment representations that disassemble large scenes into objects, regions, etc., and represent them as nodes \citep{hughes2022hydra, guConceptGraphsOpenVocabulary3D2023, werbyHierarchicalOpenVocabulary3D2024}.
The resulting representation provides a hierarchical and object-centric abstraction that has proven useful in particular in higher-level reasoning and planning \citep{rana2023sayplan, honerkampLanguageGroundedDynamicScene2024}.
In contrast to pure geometric representations, edges mainly focus on semantic or relational attributes, resorting back to grid-based maps for more detailed distance calculations.

All of the above representations establish a relation \emph{between multiple salient locations in the world}, including the agent's own position, and therefore represent knowledge about \emph{where the agent is}.

\subsubsection{Directional Cues Relative to Salient Location(s)}
On the other hand, one could imagine a spatial representation of sensory input that encodes a relation between the agent and salient locations, \emph{without knowledge about how these relate to each other}.
For example, the insect navigation base model proposes the use of view memories, which are not attached to any specific location, as discussed in more detail in Sec.
\ref{sec:spatial_rep_insect}.
One can interpret visual similarity as a proxy for the distance to the stored view and the similarity gradient as a directional cue \citep{zeilCatchmentAreasPanoramic2003} towards the location of the snapshot.
More recent models based on visual familiarity \citep{baddeleyModelAntRoute2012,ardinUsingInsectMushroom2016} allow visual homing based on stored view memories regardless of the temporal sequence or locations of the stored views.
\citet{wystrachNeuronsPremotorAreas2023} proposes a visual steering model that categorizes current views into left/right facing with regard to a specific location.
These models demonstrate that spatial representations that tell the agent \emph{where to go, rather than where it is}, are sufficient to support surprisingly complex navigation behavior.

\subsection{How is it Represented?
    Explicit and Latent Representations of Space: $\phi_{exp}$ and $\phi_{lat}$}
\label{subsec:explicit_latent_reps}
We introduce some formal definitions to pose the navigation task as a reinforcement learning problem.
Note that while we illustrate the following considerations in the context of RL, they equally apply to other learning paradigms used in the robot navigation literature.
RL is usually formalized as a Markov Decision Process (MDP)\footnote{Technically, most classical RL navigation problems are formulated as partially observable MDPs (POMDP), which operate on probabilistic belief states about the unobservable true state.
}, which is specified as a 4-tuple $(\mathcal{S}, \mathcal{A}, P, R)$:
\emph{State space} $\mathcal{S}$ and \emph{action space} $\mathcal{A}$ characterize the agent, while the environment\footnote{`Environment' in this context may also include the sensory processing apparatus} is specified by the (probabilistic) transition function
\begin{align}
    P(s^\prime | s, a), \;\;\; s,s^\prime \in \mathcal{S}, a \in \mathcal{A}
    \label{eq:transition}
\end{align}
between states $s$ and $s^\prime$, given action $a$, and a reward function
\begin{align}
    R(s,a), \;\;\; s\in\mathcal{S}, a \in\mathcal{A}.
    \label{eq:reward}
\end{align}
At each timestep, the agent moves across the state space by choosing an action, which determines the next state according to Eq.~\eqref{eq:transition}, and receives rewards according to Eq.~\eqref{eq:reward}.
The agent's objective is to learn a (probabilistic) policy
\begin{align}
    \label{eq:policy}
    \pi (a | s), \;\;\; s \in \mathcal{S}, a \in \mathcal{A}
\end{align} over actions given the agent's current state, such that under repeated applications of $\pi$, starting from any state $s$, it maximizes the expected discounted cumulative reward, which we will discuss in more detail in Sec.~\ref{subsec:RL_formalim}.
Note that our notation is meant to implicitly include policies over temporal sequences of states, e.g. eligibility traces \citep{suttonReinforcementLearningIntroduction2018}.

\begin{figure}
    \centering
    \includegraphics[width=.7\linewidth]{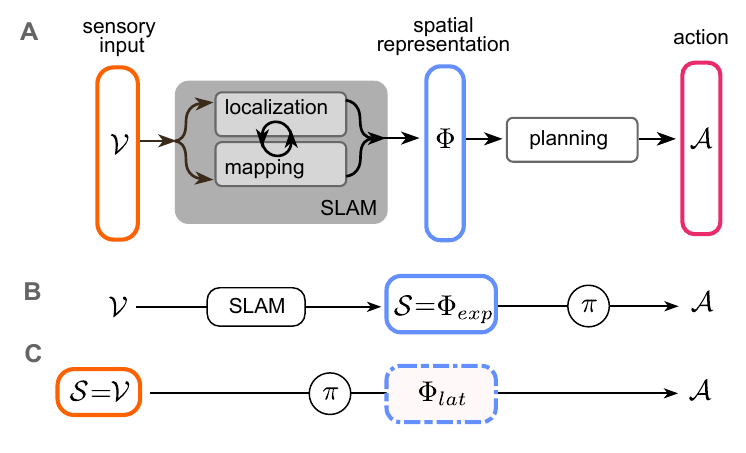}
    \caption{Navigation as a Markov Decision Process (MDP). (A) The three sub-problems of navigation: Mapping and localization operations take sensory input from an input space $\mathcal{V}$ to infer a map of the environment and the agent's pose, in a space $\Phi$ of joint spatial representations.
        These tasks are usually solved together using simultaneous localization and mapping (SLAM).
        Based on the spatial representation, the agent plans a sequence of actions from action space $\mathcal{A}$ to achieve the objective of the navigation task.
        (B) In modular robot navigation, only the planning stage is represented as an MDP: existing SLAM methods are used to construct an \emph{explicit} spatial representation $\Phi_{exp}$, which serves as the state space $\mathcal{S}$ of the MDP.
        The learned policy $\pi(\mathcal{A} | \Phi_{exp})$ then 'plans' the action.
        (C) End-to-end RL navigation directly uses the input space $\mathcal{V}$ as the MDP state space $\mathcal{S}$.
        Localization, mapping, and planning are jointly solved by learning policy $\pi(\mathcal{A}|\mathcal{V})$.
        The spatial representation $\Phi_{lat}$ is now \emph{latent} in the hidden layers of the learnable (deep) policy network.
    }
    \label{fig:NavigationMDP}
\end{figure}

In the context of a navigation task, as outlined in Sec.~\ref{subsec:geom_content}, there are now two possible choices for the state space $\mathcal{S}$ of the agent: the conventional approach was to use modular
SLAM methods to construct a spatial representation from the sensory input, and then use this \emph{explicit} representation as the state space of the agent: $\mathcal{S}\equiv \Phi_{exp}$ (see Fig.~\ref{fig:NavigationMDP}B).
The agent effectively only solves the \emph{planning} sub-problem by either planning or learning a policy over a space of spatial representations whose geometric content is pre-determined by the specific SLAM implementation (e.g. Fig.~\ref{fig:space_latent_rep}B-D)

The other possibility is to model the navigation task as an end-to-end RL (or generally end-to-end learning) problem.
This more recent approach takes the raw sensory input as the RL state space: $\mathcal{S} \equiv \mathcal{V}$ (Fig.~\ref{fig:NavigationMDP}C).
The policy learned in this case represents a joint implementation of all the sub-problems of the navigation task.
In order to achieve this, a sufficiently expressive network architecture needs to be chosen for learning $\pi$.
For example, Deep Reinforcement Learning (DRL) leverages deep neural networks as trainable function approximators (see \citealt{zhuDeepReinforcementLearning2021,zengSurveyVisualNavigation2020} for reviews of DRL for navigation tasks).
Crucially, successful end-to-end learning of a navigation task implies the existence of an implicit, or \emph{latent spatial representation} $\Phi_{lat} \neq \mathcal{S}$ within the network architecture for $\pi$
(Fig.~\ref{fig:space_latent_rep}G-I).
Notably, this representation is learned dynamically, ideally converging towards a representation that most efficiently encodes spatial features, not (fully) determined prior to training, relevant for the navigation task.
However, the space of possible representations is constrained by the network architecture, which allows imposing certain structural characteristics.

From a biological perspective, the distinction between latent and explicit spatial representation is largely a question of plasticity, recurrent connections, and time scales: In order for a latent - i.e. learned - representation to emerge, sufficient synaptic plasticity is required in the neuronal populations that encode it.
Modulation of the synaptic connections based on a training error would require some recurrent connectivity to convey that signal (see Sec.~\ref{sec:RL}).
Spatial representations in biological agents likely contain both explicit and latent components.
For example, visual processing in the insect optic lobes shows relatively little experience-dependent plasticity, but the mushroom body, which receives high-level sensory input from the visual and other sensory systems, is known as the locus of much insect learning.
From a more conceptual point of view, one could argue that the entire neuronal circuit is plastic on an evolutionary timescale.
The full spatial representation can then be interpreted as a \emph{latent representation} shaped by ecological constraints over many generations (see Sec.
\ref{subsec:evolutionary_perspective}).
While modeling explicit neuronal implementations of learning mechanisms would be nonsensical in this case, the approach may well serve as a normative model for a unified spatial representation in navigating insects.

\section{Spatial Representations in Robot Navigation}
\label{sec:spatial_rep_robo}

Having discussed \emph{what} is represented, here we discuss \emph{how} space is represented in robots.
\subsection{Explicit Spatial Representations: Variations of SLAM.}
For completeness, we will very briefly discuss traditional robot navigation models that construct explicit spatial representations using some variation of SLAM.
\citet{fuentes-pachecoVisualSimultaneousLocalization2015} provide a concise review of popular visual SLAM approaches, which operate on different modalities (monocular, stereo, multi-camera or RGB-D vision) and use different probabilistic (Extended Kalman Filter (EKF), Maximum Likelihood (ML) or Expectation Maximization (EM)) or purely geometric (`Structure from Motion') approaches to maintain a joint representation of the agent's pose within a map of the environment.
These map representations come in any of the flavors discussed above.
Egocentric occupancy grid maps \citep[][see Fig.~\ref{fig:space_latent_rep}D]{gutmann3DPerceptionEnvironment2008, xiao2022autonomous, schmalstieg2022learning} are common for dense indoor environments where obstacle avoidance is paramount.

Vector maps \citep[e.g.][see Fig.~\ref{fig:space_latent_rep}C]{kleinParallelTrackingMapping2007}, only encode the relative locations of salient features (landmarks) which are tracked in the (retinotopic) camera view across time.
Typically, these methods \citep{mur-artalORBSLAM2OpenSourceSLAM2017} use a loop of (visual) odometry based on these landmarks\footnote{The landmarks encoded in the map and the ones used for odometry need not be identical.
} and landmark prediction based on self-motion estimates to maintain a joint probabilistic estimate of robot and landmark positions.
Long-baseline feature matching allows for drift correction by loop-closure \textendash{} the recognition of previously visited locations.
In contrast to the vector memory in the INBM discussed below, this concept of landmark based maps goes beyond our earlier conceptual definition of vector maps: not only are the physical landmark locations stored in the vector map, but they are also linked to their retinotopic locations in the camera frame.

Topological SLAM methods \citep[see Fig.~\ref{fig:space_latent_rep}B]{konoligeViewbasedMaps2009,engelLSDSLAMLargeScaleDirect2014,greve2023collaborative,vodisch2022continual} are particularly useful for mapping larger areas: the world is represented as a graph in which nodes are key-frames (`sensor snapshots') representing the camera pose.
Nodes are connected by edges which represent the relationship between poses (pose-pose constraints) obtained from odometry or loop-closure.
Global optimization (e.g. Pose Graph Optimization (PGO)) ensures convergence of the topological map.
Nevertheless,
\citet{fuentes-pachecoVisualSimultaneousLocalization2015} state that due to \emph{`the lack of metric information, [\dots] it is impossible to use the map for the purpose of guiding a robot'}, a limitation which has been overcome by using latent topological representations as in \citet{shahRapidExplorationOpenWorld2023}, discussed below.

\subsection{Latent Representations.}
Recently, the attention of RNav research has shifted towards end-to-end learning approaches.
While these offer the possibility of abstract spatial representations, some implementations choose architectures that constrain the spatial representation to known templates.
In the following, we will map out (see Fig.~\ref{fig:space_latent_rep}G-I).
the space of possible latent representations along a (non-exhaustive) selection of instructive examples:

The Cognitive Mapper and Planner (CMP) model \citep{guptaCognitiveMappingPlanning2019} uses a fully differentiable encoder-decoder architecture to create a grid map of the environment.
Instead of occupancy (or pre-defined semantic) values, however, \emph{`The model learns to store inside the map whatever information is most useful for generating successful plans'}, making the map a latent representation\footnote{This is in contrast to \citep[e.g. neural SLAM][]{chaplotLEARNINGEXPLOREUSING2020}, which is explicitly trained to produce an occupancy map against a ground truth.
}.
Earlier, the same authors \citep{guptaUnifyingMapLandmark2017} suggested a latent representation that combines grid-based with vector (landmark) based maps by synthesizing a global allocentric grid-map from multiple local egocentric grid maps at salient locations.
Learning a map from egocentric observations can be viewed as storing encoded egocentric views in a map-like memory.
Explicit memory-based models like MapNet \citep{henriquesMapNetAllocentricSpatial2018} use a Long Short-Term Memory (LTSM) type Recurrent Neural Network (RNN) with convolutional layers to encode and continually update a grid-map-like state vector by egocentric observations (see Fig.~\ref{fig:space_latent_rep}H)

The RECON (Rapid Exploration Controllers for Outcome-driven Navigation) model by \citet{shahRapidExplorationOpenWorld2023} uses a network architecture whose latent representations capture the topology of a large-scale environment.
The map is represented as a graph with egocentric views (`goal images') at specific locations as nodes (see Fig.~\ref{fig:space_latent_rep}G), which are determined by a goal-directed exploration algorithm\footnote{akin to the `key-frames' in pose graphs \citep{engelLSDSLAMLargeScaleDirect2014}}.
The model employs a variation of the information bottleneck architecture \citep{alemiDeepVariationalInformation2019}: an encoder-decoder pair, conditioned on the current egocentric view - learns to compress the goal image into a latent representation (conditional encoder), which is predictive of both the (temporal) distance to the goal, and the best action to reach it (conditional decoder).
The encoder and decoder are trained together in a self-supervised manner to learn the optimal (most predictive) latent representation, with the actual time to reach the goal as ground truth.
Crucially, the resulting conditional latent representation now encodes the relative distance to the nodes, and thus the topology of the environment.
In contrast to topological SLAM models, goal-directed \emph{actions} are learned alongside the topology, enabling successful robot navigation.

Memory-based approaches like MapNet are based on the insight that all spatial representation is inherently contained in the history of previous observations.
However, these need not necessarily resort to fixed grid-like spatial representations.
Unstructured end-to-end RL approaches using RNNs \citep[e.g.]{beechingDeepReinforcementLearning2019} build completely abstract latent representations.
Similarly, the Scene Memory Transformer (SMT) architecture \citep{fangSceneMemoryTransformer2019} learns an abstract representation free of inductive biases about the memory structure.
Instead of updating an RNN state vector with each observation, an efficient embedding of every observation is stored in an unstructured scene memory.
This serves as the state space for an attention-based policy network based on the Transformer architecture \citep{vaswaniAttentionAllYou2017}, which enables the model to transform the embedding of each memory item according to a specific context.
In a nutshell, the transformer blocks are used to `[\dots] first \emph{encode} the memory by transforming each memory element in the context of all other elements.
This step has the potential to capture the spatio-temporal dependencies in the environment.~'\citep{fangSceneMemoryTransformer2019}
Thus, the encoded scene memory contains a completely abstract latent spatial representation without any preimposed structure (Fig.~\ref{fig:space_latent_rep}I).
A second attention block is then used to \emph{decode} the current observation in the context of the transformed (encoded) scene memory into a distribution over actions.
The lack of prior assumptions about spatial representation makes this model very versatile and allows applications in a variety of navigation domains.
\citet{waniMultiONBenchmarkingSemantic2020} compare models using map-based and map-less spatial representations on a multi-object navigation task.

\section{Spatial Representations of the Insect Navigation Base Model}
\label{sec:spatial_rep_insect}

In this section, we discuss \emph{how} space is represented in insects.
We describe constituent components of the proposed insect navigation base model INBM \citep{webbInternalMapsInsects2019} and analyze inherent spatial representations in the light of the previous discussion for artificial agents (see Fig.~\ref{fig:space_latent_rep}F).
Current INav research has identified three main mechanisms as the minimal set of assumptions that may be sufficient to explain observed navigation behavior.

\subsection{Path Integration}
Central place foraging insects are able to maintain a reasonably accurate estimate of their position with respect to a central nest location as a vector-like representation, known as the path integration (PI) home vector.
\citet{stoneAnatomicallyConstrainedModel2017} propose an anatomically constrained model for path integration in the central complex (CX) region of the bee brain: a self-stabilizing representation of the current heading direction is maintained in the ring-attractor architecture of the protocerebral bridge (PB): Neuronal activity of TB1 neurons in eight (per hemisphere) columnar compartments of the PB encodes heading direction relative to the sky compass, projected onto eight axes shifted by $360/8 =45 ^\circ$, leading to a periodic, sinusoidal activity pattern\footnote{This can be interpreted as a redundant - and thus more robust - generalization of simple Cartesian encoding (along two axes).}.
Another population of CX neurons (CPU4) accumulates a speed signal derived from optic flow, modulated by the current heading direction signal from the PB neurons.
As a result, the PI home vector is again (redundantly) encoded by its projection along eight axes.
This representation essentially amounts to a (discrete) phasor representation of the home vector, with the amplitude and phase of the periodic signal representing its length and angle, respectively.
The home vector can then be used to drive the animal back towards the nest.
In the context of this work, we want to stress two important aspects of PI in flying navigators: First, it must rely to a large extent on vision alone, since proprioceptive modalities used by walking insects like desert ants are highly unreliable due to wind drift and other atmospheric parameters.
We can therefore interpret the PI home vector as a predominantly visual representation of space\footnote{The role of vestibular-like inertial sensory information in PI is largely unknown.
}.
Secondly, for the same reason, basing PI on \emph{heading direction} is an oversimplification since heading and traveling direction will often differ.
\citet{lyuBuildingAllocentricTravelling2022} proposed a circuit model of the fly (\emph{Drosophila melanogaster}) CX, demonstrating how a representation of the \emph{allocentric traveling direction} can be computed from heading direction and \emph{egocentric directional} optic flow cues by a phasor-based neural implementation of vector addition.
However, this important implementation detail does not invalidate the general PI mechanism outlined above.

\subsection{Vector Memory}
Efficient navigation entails more than returning to the nest: Foragers need to be able to reliably revisit known food sources.
The INBM posits that whenever insects visit a salient location, they store the current state of the home vector in a vector memory.
\citet{lemoelCentralComplexPotential2019} suggest a mechanism where an individual vector memory is stored in the synaptic weights of a memory neuron, which forms tangential inhibitory synapses onto all directional compartments of the CPU4 population.
Activation of this neuron when the home vector is zero would leave a negative imprint of the memorized PI vector (i.e. the vector from the nest to the remembered location) in CPU4 activity, driving the animal to recover a CPU4 activity corresponding to a zero vector (which is now the case at the remembered location, where vector memory and home vector are equal).
If the initial home vector is not zero, this mechanism for vector addition effectively computes the direct shortcut to the remembered location, an ability frequently cited as strong evidence for the existence of a cognitive map\footnote{Another implementation of a vector memory may be sustained neuronal `phasor type' activity of (unknown) cell types in the CBU, and performing vector arithmetic as in \citet{lyuBuildingAllocentricTravelling2022}
}.
The neuronal mechanisms for storing, retrieving, and choosing between multiple vector memories remain speculative.
For the former, the authors suggest dopaminergic synaptic modulation directly at the CPU4 dendrites, providing direct reinforcement from extrinsic rewards (food).

This combination of accurate path integration and vector memories constitutes a vector map, i.e. a geometrically accurate (within limits of PI accuracy) representation of the world, which the insect can access for navigation, as long as the PI vector is not corrupted or manipulated.
Unlike landmark-based maps in vSLAM, the vector locations are not associated with any visual landmarks or features.

\subsection{View Memory}
A large body of INav research has been concerned with the ability to return the nest when an accurate PI vector is not accessible to the animal, making it reliant on visual homing and route-following mechanisms.
Originating from the snapshot model \citep{cartwrightLandmarkLearningBees1983} which matched the retinal positions of landmarks between the current view and stored snapshots, more recent models suggest retinotopic representations of a low-resolution panoramic view, with only elementary processing like edge filters, and without the need for explicit landmark recognition.
\citet{zeilCatchmentAreasPanoramic2003} showed that similarity gradients based on pixel-by-pixel intensity differences are sufficient for successful visual homing.
Combining multiple view memories along frequently traveled routes allows for complex routes following toward the nest.
\citet{webbInternalMapsInsects2019} emphasizes that no information about the location or temporal sequence of the stored views is necessary: \citet{baddeleyModelAntRoute2012} proposed a computational familiarity model, which encodes the entire view memory in an InfoMax \citep{leeIndependentComponentAnalysis1999,lulhamInfomaxAlgorithmCan2011} neural network architecture.
From a scan of the environment, the agent can then infer the most familiar viewing direction over all stored memories.
If the view memories are acquired during inbound routes (i.e. linked to a homing motivational state), this will guide the agent towards the nest.
\citet{ardinUsingInsectMushroom2016} proposed a biological implementation of a familiarity model based on the insect mushroom body (MB), a learning-associated region of the insect brain discussed in more detail in Sec.~\ref{sec:RL}.
In RL, different neural network approaches to assess view familiarity have been used e.g. in Random Network Distillation \citep{burdaExplorationRandomNetwork2018}.
Instead of a homing signal, it serves as motivation for exploring unknown states (cf.
discussion Sec. \ref{sec:discussion})

\subsection{Discussion: Spatial Representation in the Insect Navigation Base Model}
As presented thus far, the base model entails two independent spatial representations: A vector map, which is not linked to specific egocentric views, and directional cues based on view memories, which are not linked to any geometric information from the vector memory (see Fig.~\ref{fig:space_latent_rep}F).
According to our previous classification, the vector map is an \emph{explicit spatial representation}: It is evolutionarily pre-determined by the path integration circuitry, just like classical SLAM architecture by a pre-defined inference algorithm.
The construction of the vector representation differs from SLAM methods in that exact localization is assumed, based on PI, and the map is constructed based on that ground truth, obviating the need to maintain correspondences between retinotopic and geometric locations of features and landmarks.
On the other hand, a crude \emph{latent spatial representation} in terms of directional cues is implicit in the view memory.

For example, the visual features used by the InfoMax architecture for familiarity discrimination are latent in the learned network weights.
Fig.~\ref{fig:space_latent_rep}F illustrates how the INBM aligns with our classification of spatial representations used for robot navigation.
As mentioned, the base model explicitly does not link view memories to vector memories.
\subsection{Beyond the base model.}
How these two distinct representations are linked is an active research question, covering two major aspects: First, how do insects balance conflicting information from the two systems?
\citet{sunDecentralisedNeuralModel2020} proposed a unified model inspired by joint MB/CX neuroanatomy, combining PI, visual homing, and visual route following.
The model balances off-route (PI and visual homing) with on-route (visual route following) steering outputs based on visual novelty and uncertainty of the PI signal.
\citet{goulardEmergentSpatialGoals2023} recently proposed a different mechanism to integrate view based and vector based navigation, based on an extended concept of vector memories: Positing vector computations as the fundamental scaffold for integrating navigational strategies, they exploit the fact that both PI home vector and the current heading direction are represented in ring attractor networks within the CX \citep[c.f.][]{stoneAnatomicallyConstrainedModel2017,lemoelCentralComplexPotential2019}.
In addition to \emph{location specific} vector memories (i.e. imprints of the home vector, as in \citet{lemoelCentralComplexPotential2019}) they introduce \emph{direction specific} memories, which store the current heading direction via inhibition by a second set of `memory neurons'.
View familiarity is again assumed to be learned in the MB, producing a binary signal which is used to store the instantaneous heading direction as a vector memory, whenever a familiar view is encountered.
View familiarity is thus translated directly into a directional vector memory, which can be combined with the PI based home vector (or other location specific vector memories) to produce a final steering output.

Conceptually more interesting is a second aspect: is the view memory truly independent of the geometry of the vector memory?
This is closely related to a question not thoroughly addressed in the work cited above: When and where does an animal form a view memory?
Most models just assume that views are stored regularly along a homeward-bound route.
\citet{ardinUsingInsectMushroom2016} suggest that \emph{`the home reinforcement signal could [\dots] be generated by decreases in home vector length'}.
Note that this already associates the stored views with a specific node in the vector map.
\citet{wystrachNeuronsPremotorAreas2023} recently proposed a neuroanatomically constrained model for visual homing which obviates the need for storing individual view memories: during learning, views are continuously associated with facing left or right with respect to the nest, using the difference between PI vector and current heading direction for reinforcement.
We will discuss this model in detail in Sec.~\ref{subsec:visualhoming}.
The spatial representation of egocentric views is now decidedly conditioned on a specific vector.
One could easily imagine an extension of this model by vector memories, enabling learnable visual guidance along arbitrary vectors encoded in the vector map, essentially using each vector as a motivational state.
Note that such a mechanism would be different from `reloading' a PI state from the view memory, although the expected behavior is similar: Insects would be able to recover previously known `shortcuts' based on visual guidance alone.
The joint spatial representation would be a topological latent representation similar to \citet{shahRapidExplorationOpenWorld2023}, see the discussion in Sec.
\ref{sec:discussion}.

\subsection{An Evolutionary Perspective: Insect Inspired RL as a Normative Model}
\label{subsec:evolutionary_perspective}
Conceptually, such a unified spatial representation could itself be viewed as a single, latent embedding of visual input, learned over evolutionary time to best adapt to ecological constraints, i.e. reap the largest long-term reward from the environment.
The dichotomy between static, explicit versus plastic, latent components of the representation would then be relaxed to a continuum of plasticity for different model components, realized via differential learning rates.
We propose to design an end-to-end RL-learnable navigation model constrained by the insect navigation base model, in the sense that the resulting spatial representation is compatible with its basic assumptions.
This will be instructive for both the field of insect and robot navigation: For the former, it can serve as a normative model for a possible unified spatial representation that goes beyond the base model, providing theoretical guidance for how vector and view-based representations may interact to support efficient navigation.

On the level of sensory processing, having the network learn representations that match, e.g., PI based vector maps may yield valuable insights into which visual features are useful in intermediate processing steps to reliably support such computations in a variety of visual conditions and environments.
These model predictions would yield testable hypotheses for further neuroanatomical, physiological, and behavioral experiments, as discussed in Sec.~\ref{sec:discussion}.
Given the superior performance of insect navigators in terms of training efficiency, robustness, and generalization capability, robot navigation may profit from this biologically inspired and constrained spatial representation.
Pretraining such a network extensively under varying conditions and then freezing the slow components may yield a highly robust, adaptive spatial representation for applications similar to natural insect task spaces, e.g. visual outdoor navigation for ground or aerial autonomous agents.
Implementing network architectures that support phasor representations may be a useful avenue for robotic navigation research.

\section{Reinforcement Learning with an Insect Brain}
\label{sec:RL}

Given the success of reinforcement learning as a framework for robot navigation, it seems reasonable to ask if and how navigation could be implemented based on actual RL-type computations in the insect brain.
Furthermore, extensive literature involving dopamine, learning, and reward prediction errors exists in the mammalian neuroscience community but despite these topics being relevant in insect learning and navigation, the discussion of potential connections is limited.
To explore this line of thought, we will first continue our formal treatment of RL (Sec.
\ref{subsec:RL_formalim}).
In Sec.~\ref{subsec:MB_canonical} we will discuss how current computational models of the MB - the prominent learning associated region of the insect brain - could be augmented to support simple RL algorithms.
This will allow us to discuss the recent MB/CX based visual homing model by \citet{wystrachNeuronsPremotorAreas2023} in the context of RL and extrapolate it to roughly outline a neuroanatomically inspired end-to-end RL model for insect navigation in Sec.~\ref{sec:RL_insect_navigation}

\subsection{RL Formalism}
\label{subsec:RL_formalim}
Starting from the definitions from Sec.~\ref{subsec:explicit_latent_reps}, RL methods find the optimal policy Eq.~\eqref{eq:policy} which may maximize the expectation of the temporally discounted sum of instantaneous rewards Eq.~\eqref{eq:reward} over time:
\begin{align}
    V^\pi(s) & = \mathbb{E}_\pi\left[\sum_{k=0}^{T} \gamma^k R_{t+k+1} \, \middle| \, S_t=s \right] \\ & = \mathbb{E}_\pi \left[ R_{t+1} + \gamma V^\pi (S_{t+1} ) \, \middle | \, S_t = s \right] \label{eq:BellmanV1} \\ & = \sum_a \pi (a|s)\left[ R(s,a) + \gamma \sum_{s^\prime} p( s^\prime | s, a) V^\pi(s^\prime) \right] \label{eq:BellmanV2} \end{align} for all initial states $s$.
This function is therefore called the \emph{value function} of $s$ under policy $\pi$, with a temporal discount factor $\gamma \in [0,1]$.
Eq.~\eqref{eq:BellmanV1} and Eq.~\eqref{eq:BellmanV2}, versions of the \emph{Bellman equation}, illustrate the recursive nature of the value function: it can be decomposed into the average immediate reward from the current state $s$ under policy $\pi$, plus the discounted value of the subsequent state, averaged over all possible successor states $s^\prime$.
Maximization of $V$ with respect to $\pi$ can now be understood intuitively: For a single time step, the optimal policy $\pi^*$ for the recursion Eq.~\eqref{eq:BellmanV2} would be simply choosing the action which maximizes the term in square brackets.
Iterating through the recursion then leads to the \emph{Bellman optimality equation} for the optimal state value function
\begin{align}
    V^*(s) & = \max_a \left[ R(s,a) + \gamma \sum_{s^\prime} p( s^\prime | s, a) V^*(s^\prime) \right]
    \\
           & = \max_a \mathbb{E}\left[ R_{t+1} + \gamma  V^*(S_{t+1}) \,\middle |\, S_t=s, A_t=a \right]
    \label{eq:BellmanOptV}
\end{align}
Another way to interpret Eq.~\eqref{eq:BellmanV2} would be as a policy average over a \emph{state-action value function} $Q^\pi(s,a)$:
\begin{align}
    V^\pi (s) \equiv \sum_a \pi (a|s) \, Q^\pi(s,a).
\end{align}
By the same logic, recursive (optimality) relations can be derived for $Q$:
\begin{align}
    Q^\pi(s,a) & = \mathbb{E}_\pi \left[ R_{t+1} + \gamma Q^\pi(S_{t+1}, A_{t+1}) \, \middle | \, S_t=s, A_t=a \right] \label{eq:BellmanQ} \\ Q^* (s,a) & = \mathbb{E} \left[ R_{t+1} + \gamma \max_{a^\prime} Q^*(S_{t+1}, a^\prime) \, \middle | \, S_t=s, A_t=a \right] \label{eq:BellmanOptQ} \end{align}

\subsubsection{Value-based, Policy-based, and Actor-Critic Methods}

One way to find the optimal policy $\pi^*$ is by trying to solve it directly.
This is most commonly done using policy gradient methods which parameterize the policy $\pi(a|s; \theta)$ and then perform gradient ascent on a suitable performance metric, like the average reward per timestep: $\Delta \theta \propto \nabla_\theta \mathbb{E}[R_t]$.

We will not dwell on pure policy-based methods further, for more detail see \cite{suttonPolicyGradientMethods1999, williamsSimpleStatisticalGradientfollowing1992}.
Alternatively, one can estimate the value function $V(s)$ or $Q(s,a)$ and infer an optimal policy indirectly.
For example, assuming an optimal $Q^*$ is found, the optimal policy is simply
\begin{align}
    \pi^* = \text{argmax}_a Q^*(s,a)
    \label{eq:greedy_argmax}
\end{align}
These value-based approaches have the strongest connection to insect neuroscience and will therefore feature prominently in the rest of this paper.
\\
Last, so-called actor-critic methods constitute a hybrid approach, involving both a policy (actor) and value (critic) estimation.
The policy gradient is then computed to maximize the \emph{advantage} of an action derived from the policy over the estimated baseline value.
Actor-critic methods play an important part in robot navigation.

\subsubsection{Temporal Difference (TD) Methods: SARSA and Q-learning}
TD methods approach the problem of value estimation by deriving single-timestep update rules from the recursive relations Eq.~\eqref{eq:BellmanQ}, Eq.~\eqref{eq:BellmanOptQ} for $Q$ or Eq.~\eqref{eq:BellmanV1}, Eq.~\eqref{eq:BellmanOptV} for $V$: for each timestep, the squared difference between the LHS and RHS is treated as a prediction error to be minimized.
For Eq.~\eqref{eq:BellmanQ}, this leads to the update rule
\begin{subequations}
    \label{eq:SARSA}
    \begin{align}
        Q^\pi(S_t,A_t) & \leftarrow Q^\pi(S_t,A_t) + \alpha \cdot\Delta_t, \label{eq:SARSA_update}                   \\
        \Delta_t =     & \left[R_{t+1} + \gamma Q^\pi(S_{t+1}, A_{t+1}) - Q^\pi(S_t,A_t) \right] \label{eq:SARSA_TD}
    \end{align}
\end{subequations}
with learning rate $\alpha \in [0,1]$.
This update rule is known as SARSA due to the tuple $(S_t,A_t,R_{t+1}, S_{t+1}, A_{t+1})$ required to compute the update.
In order to accommodate exploration, the agent follows a non-deterministic policy based on the current estimate of $Q$, making SARSA an \emph{on-policy} algorithm.
The most common choices are $\epsilon$-greedy (choose the $\text{argmax}_a(Q)$ with probability $1-\epsilon$, random action with probability $\epsilon$) and softmax policies (choose an action from a Boltzmann distribution based on $Q$, with inverse temperature $\beta$).
These policies converge to the optimal policy $\pi^*$, if the stochasticity is reduced systematically towards a deterministic (greedy) policy ($\epsilon \rightarrow 0$ or $\beta \rightarrow \infty$) during learning \citep[see][]{suttonReinforcementLearningIntroduction2018}.

Finally, a TD error without reference to a specific policy can be derived to estimate $Q^*$ directly from Eq.~\eqref{eq:BellmanOptQ}:
\begin{align}
    \Delta_t =  \left[R_{t+1} + \gamma \max_a Q^*(S_{t+1}, a) - Q^*(S_t,A_t) \right]
    \label{eq:Q-learning}
\end{align}
This is known as Q-learning.
Note that in contrast to Eq.~\eqref{eq:SARSA_TD}, this \emph{off-policy} update is now independent of the consecutive action $A_{t+1}$ prescribed by the policy.
However, the policy still determines which action-value pair receives the update.
This effectively decouples the learned policy from the policy employed during learning.
In particular, this also allows for the agent to perform \emph{off-line} updates, i.e. updates which are not based on the current state transition, but e.g. sampled from a replay buffer of previous experiences.
This property was key to the success of the Deep Q-Network (DQN) by \citet{mnihHumanlevelControlDeep2015}, an early milestone of Deep Reinforcement Learning, which has since found numerous applications in robot navigation.

As we will show, the neural substrate of the insect mushroom body has the potential to support $Q$-based TD computations like SARSA or Q-learning to solve navigational tasks.
Conversely, (Deep) Reinforcement Learning as a key toolset for robot navigation provides a useful framework to think about the neural computations underlying insect navigation.

\subsection{The Mushroom Bodies as a Neural Substrate for RL}
\label{subsec:MB_canonical}
\subsubsection{The canonical MB learning model for classical conditioning.}
\label{subsubsec:MB_canonical}

\begin{figure*}
    \centering
    \includegraphics[width=0.89\textwidth]{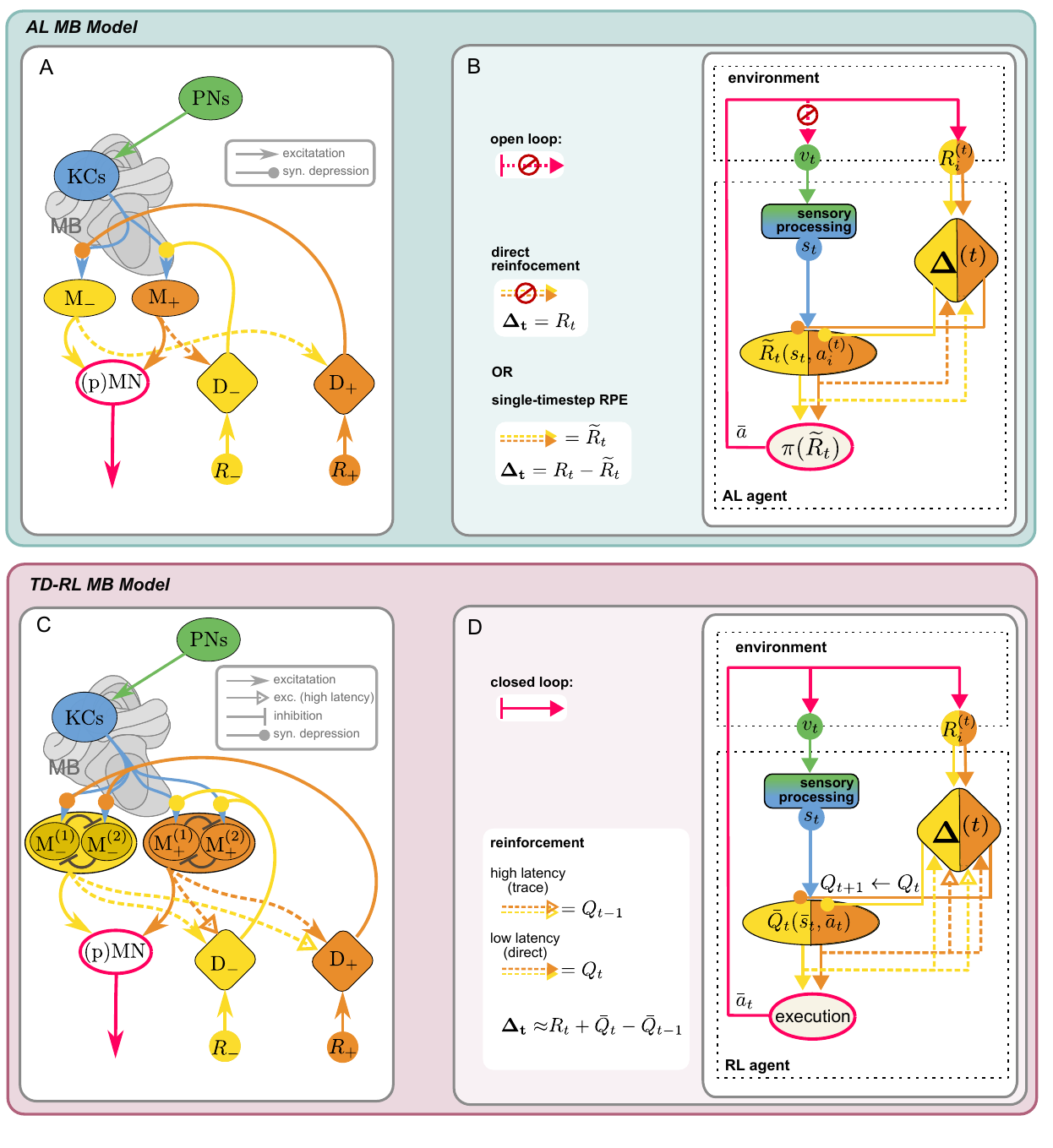}
    \caption{Anatomical (A, adapted from \citealt{bennettLearningReinforcementPrediction2021} licensed under \href{https://creativecommons.org/licenses/by/4.0/}{CC-BY 4.0}) and computational (B) components of MB model for AL: KCs receive sensory input $v_t$ from the PNs, encoding a decorrelated and sparse representation of the sensory environment $s_t$.
        KCs synapse onto distinct sets of approach/avoid MBONs ($M_\pm$) driving opposing responses, which are integrated in downstream (pre)motor neurons, $\text{(p)MN}$, to produce an effective action, according to a `policy' $\pi$.
        MBON activity can be interpreted as a prediction $\widetilde{R}$ of external reward $R$, stored in KC$\rightarrow$MBON synaptic weights.
        These are depressively modulated by reinforcement signals $\Delta_\pm$ from distinct sets of aversive/appetitive ($D_\mp$) DANs of \emph{opposite} valence.
        In the canonical model, DANs encode direct (external) positive or negative reinforcement $\Delta = R$.
        Including recurrent MBON $\rightarrow$ DAN connections (dashed) enables DANs to compute an RPE signal for reinforcement: $\Delta = R - \widetilde{R}$ (see Sec.~\ref{subsec:MB_canonical}).
        (C,D) MB as a neural substrate for RL (Sec. \ref{subsubsec:MB_RL}):
        When the state-action loop is closed, $s_t$ are valid states of an MDP.
        A discrete action space (here: $n=2$) is represented by corresponding sets of MBONs (in `conjugate' valence pairs).
        A value-based policy could be implemented by lateral inhibition between MBONs (schematic illustration only).
        The MB circuit can support TD-RL with the additional assumption that high-latency recurrent MBON$\rightarrow$DAN connections can carry an activity trace.
        DAN activity can then encode a TD error \eqref{eq:TD_MB} and
        MBON activity reflects the agent's current estimate of $\bar{Q}$, the value prediction for the current state-action pair $(\bar{s}_t,
            \bar{a}_t)$.
    }
    \label{fig:MB_RL1}
\end{figure*}

The mushroom bodies are bilateral neuropils in the insect brain, with homologous structures largely conserved across different species, whose crucial role in learning and memory has long been established.
Extensively studied in the context of olfactory learning in \emph{Drosophila melanogaster} \citep[reviewed in][]{cogniniFlyLearning2018}, inputs from other sensory modalities, in particular vision \citep[e.g.][]{ehmerSegregationVisualInput2002,strubeblossMultimodal2018,vogtFlyVision2014} likely support general behavioral learning tasks, including visual-spatial navigation.
The main intrinsic anatomical components of the MB are \emph{Kenyon cells} (KC), whose dendrites form the calyx, while the axons constitute the lobes of the MB.
They receive sensory inputs via \emph{projection neurons} (PN), which are thought to form non-plastic, sparse and random \citep{caronRandom2013} synapses onto the KCs at the calyx.
KC activity is transmitted to \emph{mushroom body output neurons} (MBONs) at the MB lobes.
\emph{Dopaminergic neurons} (DANs), which also target the MB lobes, induce (usually depressive) modulation of the KC-MBON synapses and thus enable an adaptive response to the sensory stimulus.
MBON activity is integrated downstream by (pre)motor neurons (MN) to produce an action.
Conventionally, the reinforcement signal mediated by the DANs is assumed to encode a direct extrinsic reward.
Fig.~\ref{fig:MB_RL1}A shows the `canonical' MB circuitry (without the dashed MBON$\rightarrow$DAN synapses).

\paragraph{Prediction Targets and Reinforcement Signals: RL vs.
    Associative Learning.
}

At first glance, all the ingredients for reinforcement learning seem to be there: KC activity defines the state space $\mathcal{S}$, based on which MBON activity encodes \emph{some} value prediction over an action space $\mathcal{A}$.
DAN activity encodes a reward function $R(s,a)$.
However, there is a crucial difference in the \emph{kind of value prediction} which is computed: So far, MB-based learning has been studied in the context of trial-by-trial classical conditioning, a form of associative learning (AL) where the agent is presented with isolated stimulus-reward pairs.
This is \emph{not} a full MDP, since neither future states nor rewards are contingent upon the current state and action, i.e. the transition function $P(s^\prime|s, a)$ is not specified.
The prediction target of RL, the long-term cumulative reward, is therefore ill-defined, since it is determined by the experimenter's choice of stimulus/reward pairs, and not (only) by the agent's action.
Instead, the agent is trained to predict the immediate external reward following an action.
This highlights a fundamental difference in the interpretation of rewards in AL vs.
reinforcement learning: In the former, it serves as an immediate feedback signal used to evaluate individual actions.
The agent does not learn to maximize future rewards,
but merely to react to a stimulus according to the associated
reward.
In the latter, rewards define a long-term objective which
the agent learns to achieve by a series of optimal actions.

Furthermore, the models differ in \emph{how the value prediction is learned}: in the canonical MB model, the DAN reinforcement signal \emph{directly} encodes the absolute value of the external reward $R$ (direct reinforcement), while for TD-RL methods, $\Delta_t$ in Eq.~\eqref{eq:SARSA} and Eq.~\eqref{eq:Q-learning} encode a \emph{prediction error} of $Q(s,a)$ (which is a proxy for the prediction target $V(s)$).
Crucially, an MDP cannot be learned using direct reinforcement, since there is no directly provided ground truth for the prediction target.
A neural mechanism for computing prediction errors is therefore a prerequisite to reconcile RL with MB based computations.

\subsubsection{Prediction Errors in the MB.}
\label{subsubsec:MB_RPE}
Recent computational studies proposed rate-based \cite{bennettLearningReinforcementPrediction2021} and spiking \cite{jurgensenPredictionErrorDrives2024} models of the MB which employ recurrent MBON$\rightarrow$DAN connections to compute a reward prediction error (RPE).
They show that the resulting behavior of the agents in a classic conditioning paradigm aligns with experimental evidence.
The postulated recurrent connections are supported by anatomical evidence \citep[see references 17-20 in][]{bennettLearningReinforcementPrediction2021} and are indicated by the dashed arrows in Fig.~\ref{fig:MB_RL1}A.
It illustrates the simplest iteration of the models investigated by \citet{bennettLearningReinforcementPrediction2021}: Agent behavior is the net result of approach and avoid opponent processes within the MB.
The two antagonistic behaviors are encoded by the activity of two distinct sets of MBONs ($M_\pm$), whose outputs are integrated by downstream descending neurons.
KC$\rightarrow$MBON synapses of these distinct sets are targeted by two sets of valence specific, i.e. appetitive and aversive, DANs ($D_\pm$).
The major innovation of this model lies in the interpretation of $M_\pm$ activity as predictions $\widetilde{R}_\pm$ of the positive and negative external reward $R_\pm$, respectively\footnote{
    Somewhat confusingly, rewards of opposite valence are often termed `reward' and `punishment' in the AL literature.
}
The prediction error $\Delta R$ of the total reward $R = R_+ - R_-$ is then computed indirectly by recurrent excitatory MBON $\rightarrow$ DAN connections of opposite valence, i.e. the prediction of negative reward is \emph{added} to the direct positive reward, and vice versa: $D_\pm \leftrightarrow R_\pm + \widetilde{R}_\mp$.
The difference between $D_+$ and $D_-$ activity then encodes the full RPE:
\begin{align}
    \label{eq:RPE}
    D = D_+ - D_- \;\;\;\leftrightarrow\;\;\;
    (R_+ + \widetilde{R}_-) - (R_- + \widetilde{R}_+) = \Delta R_+ - \Delta R_- = \Delta R
\end{align}
Since the synaptic modulation by DANs is depressive, reinforcement is achieved by inhibiting MBONs of the opposite valence, i.e. appetitive DANs inhibit aversive MBONs and vice versa.
In contrast to the temporal difference errors from SARSA Eq.~\eqref{eq:SARSA} and Q-learning Eq.~\eqref{eq:Q-learning}, here the RPE reflects the prediction error of the single-timestep (or `timeless') total external reward of the Rescorla-Wager type \citep{rescorla1972theory}.
We propose that only a few additional assumption can turn this MB circuit into a neural implementation of a TD RL agent.

\subsubsection{Temporal Dynamics of Recurrent Connections in the MB can Support Temporal Difference Learning.}
\label{subsubsec:MB_RL}
Once we change the experimental paradigm to an interaction task that can be modeled as a full MDP, optimizing cumulative rewards becomes a meaningful objective.
The agent's learning objective is now no longer the immediate reward $R(s,a)$ following an action, but the long-term value of a state-action pair $Q(s,a)$.
The recursive TD update rules \eqref{eq:SARSA},\eqref{eq:Q-learning} are prediction errors of the current estimate of $Q(s,a)$, with a more complex prediction target: $R(s,a)$ \emph{plus} the agent's estimate for $Q(s^\prime, a^\prime)$ in the \emph{next} timestep.
This temporal link is the core principle allowing TD methods to compute estimates of cumulative rewards over time.
As before, we can interpret the activity of distinct sets of MBONs as predictions, now for $Q$ instead of $R$.

To illustrate the key differences to the RPE model discussed above, let us consider the simple case of SARSA learning on a discrete $n$-dimensionsal action space $\{a_i\}$ beyond binary approach/avoid behavior - representing for example the choice of a navigational sub-goal (`vector memory', see Sec.
\ref{subsec:MBCXRL}).
Unlike in standard `computational' SARSA, positive and negative rewards in the MB model are encoded in valence-specific pathways, leading to a valence-specific decomposition $Q = Q_{+} - Q_{-}$, represented by the activity of two distinct sets of $n$ MBONs: $M_\pm^{(i)} \leftrightarrow Q_{\pm}(s,a_i)$.
In order to implement the SARSA update rule via dopaminergic modulation, the following differences to the RPE model are crucial:
(i) To convey estimates of $Q$ at subsequent timesteps, recurrent MBON$\rightarrow$DAN connections would have to carry \emph{acitivty traces}, for example via multiple pathways with different temporal dynamics\footnote{Technically, Eq.~\eqref{eq:SARSA} looks forward in time, while an activity trace only gives access to previous states.
    This could be reconciled by simply shifting the time index such that $Q_(S_{t-1}, A_{t-t})$ is updated at time $t$.
    In the neural model, this could be captured by the temporal dynamics of synaptic modulation.
}, synaptic strength and potentially different effective signs. \citet{eschbachRecurrentArchitectureAdaptive2020} have demonstrated the existence of these different kinds of pathways in larval \emph{Drosophila} in the context of classical conditioning.
It would be interesting to investigate whether they converge in a way that enables the computation of TD errors.
(ii) The TD-error \eqref{eq:SARSA_TD} only involves predictions of $Q(\bar{s},\bar{a})$ corresponding to the actually experienced state-action pair $(\bar{s},\bar{a})$, chosen according to a value-based (probabilistic) policy: $\bar{a} \sim \pi(Q(\bar{s},a_i))$ (see Sec.
\ref{subsec:RL_formalim}).
Since MBON population activity $\{M_\pm^{(i)}\}$ encodes the entire \emph{function} $Q({\bar{s},a_i})$, this would require \emph{selective}
MBON$\rightarrow$DAN feedback, reflecting the implementation of $\pi$.
(iii) Finally, synaptic modulation must selectively affect only KC$\rightarrow$MBON synapses corresponding to $Q(\bar{s},\bar{a})$, according to \eqref{eq:SARSA_update}.

\emph{Action selection and policy implementation.}
The simplest way to implement a policy-dependent action selection is directly at the level of MBON activity.
For example, a winner-take-all type lateral inhibition mechanism between MBONs would correspond to a (greedy) argmax policy (cf.
Eq.
\eqref{eq:greedy_argmax}).
Accounting for noise in the circuit, this could be interpreted as a non-deterministic policy\footnote{This could even account for adaptive exploration, with an increasingly exploitative policy as the signal-to-noise ratio increases, i.e. the agent becomes more evaluation of a state-action pair becomes more informative (larger in absolute value)}, potentially giving rise to an $\epsilon$-greedy or softmax-like selection mechanism.
As a result, only the `conjugate' MBON pair $\bar{M_\pm}$ would be active, reflecting the value $\bar{Q}_\pm \equiv Q_{\pm}(\bar{s},\bar{a})$ of the chosen (experienced) action in the given state.
This ensures that \emph{unspecific} recurrent MBON$\rightarrow$DAN connections would still convey only the relevant $Q$-prediction such that joint DAN activity represents the TD error \eqref{eq:SARSA_TD}: Extrapolating from the RPE computation
\eqref{eq:RPE}, this could be implemented by assuming \emph{low-latency ($t$), excitatory} recurrent connections between DANs and MBONs of \emph{equal valence} and  \emph{high-latency ($t^\prime$), excitatory} connections between \emph{opposite valence} compartments:
\begin{align}
    \label{eq:TD_MB}
    \begin{split}
        D = D_+ - D_- \;\;\;\leftrightarrow\;\;\;\; & \left[R_+(t) + \bar{Q}_+(t) + \bar{Q}_-(t^\prime) \right] - \left[ + \leftrightarrow - \right] \\
        =                                           & \left[R_+(t) + \bar{Q}_+(t) - \bar{Q}_+(t^\prime) \right] - \left[ + \leftrightarrow - \right] \\
        =                                           & \Delta_{t^\prime}^+ - \Delta_{t^\prime}^- = \Delta_{t^\prime}
    \end{split}
\end{align}
If we additionally assume a DAN gated 3-factor plasticity rule at the KC$\rightarrow$MBON synapse, depending on both presynaptic KC \emph{and} postsynaptic MBON activity, updates to the value prediction would apply selectively only to $Q(\bar{s},\bar{a})$, see (ii) above.
Both lateral MBON inhibition \citep[e.g.][]{huertaLearningClassificationOlfactory2004} and effective 3-factor plasticity rules \citep[e.g.][]{huertaFastRobustLearning2009,faghihiComputationalModelConditioning2017} have been employed in MB learning models, summarized nicely in a recent review by \citet{webbPredictionError252024}.
Other combinations of mechanisms presented therein may be equally viable to effectively perform RL with a MB inspired circuit.
Alternatively, action selection may take place downstream of the MB, for example in the CX which has been suggested as a potential substrate for action selection in \emph{Drosophila} \citep{hulseConnectomeDrosophilaCentral2021}.
The recurrent MBON$\rightarrow$DAN connections described by \citet{eschbachRecurrentArchitectureAdaptive2020} also include multi-synaptic pathways, which would be necessary for such a more indirect action-selection and reinforcement mechanism.

In any case, the agent's current estimate of the \emph{function} $Q$ is fully captured by the KC$\rightarrow$ MBON synaptic weights, learned by synaptic modulation through the DANs.
Fig.~\ref{fig:MB_RL1} illustrates how a full TD model can be obtained by augmenting the canonical classical conditioning model of the MB, and how anatomical and algorithmic components map onto each other.
If rewards are given only externally at sparse locations, successful learning in such an RL agent will require training over many episodes.

\begin{figure*}
    \centering
    \includegraphics[width=1.0\textwidth]{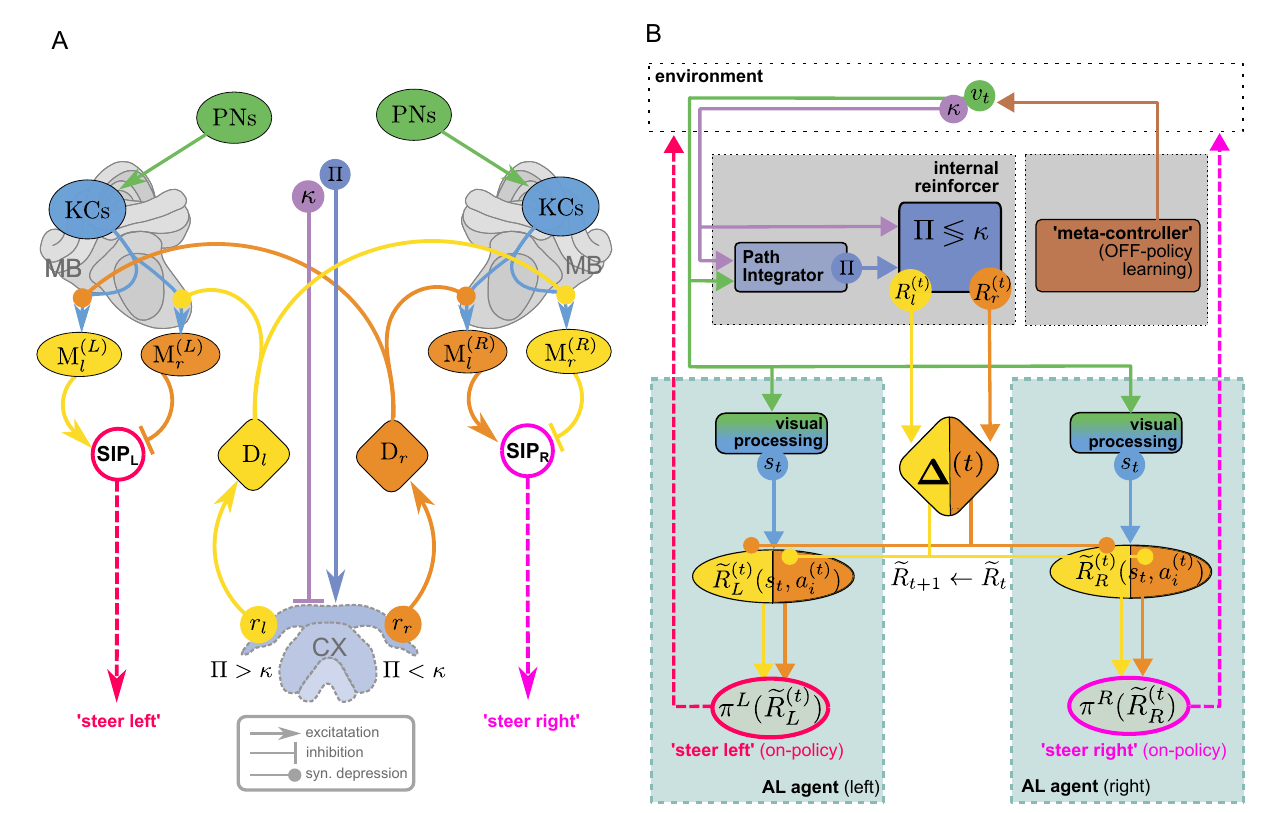}
    \caption{
        (A) The MB/CX model for vision-based homing (adapted from \citealt{wystrachNeuronsPremotorAreas2023}, licensed under \href{https://creativecommons.org/licenses/by-nc/4.0/}{CC-BY-NC 4.0}) can be viewed as two competing MB AL agents (Fig.~\ref{fig:MB_RL1}
        A,B) in the left ($L$) and right $(R)$ hemisphere, receiving internal rewards provided by vector computations in the CX (Panel B).
        Instead of positive/negative external rewards, the set of DANs now encodes whether the nest is located to the left or to the right ($D_{l/r}$) as a direct reinforcement signal.
        No RPE computation is assumed in this model.
        This `ground truth' is computed in the CX based on the current compass heading $\kappa$ and a PI home vector (whose compass angle we denote as $\Pi$, such that $\Pi < \kappa$ means that the nest is located to the left).
        In each hemisphere, competition between `steer left' and `steer right' MBONs ($M_{l/r}$) is integrated ipsilaterally by neurons in the Superior Intermediate Protocerebrum (SIP) to generate a steer left/right command in the left/right hemibrain, respectively.
        Opposing valence of the steering commands between hemispheres is achieved by inverting excitation/inhibition of the SIP inputs.
        Note that visual steering is learned `off-policy', i.e. the agent does not use the policy it learns during learning but is instead driven by an `off-policy' metacontroller.
        (In \citealt{wystrachNeuronsPremotorAreas2023} the agent was simply made to retrace experimental trajectories of learning walks.)
    }
    \label{fig:MB_RL2}
\end{figure*}

\section{Reinforcement Learning for Insect Navigation}
\label{sec:RL_insect_navigation}

\subsection{Visual Homing with Vector-based Internal Rewards}
\label{subsec:visualhoming}
\cite{wystrachNeuronsPremotorAreas2023} proposed an MB/CX based visual homing model that uses an internal reward signal to alleviate this problem.
Fig.~\ref{fig:MB_RL2}A illustrates how the model consists of two antagonistic copies (one in each hemisphere) of the `canonical' MB circuit in Fig.~\ref{fig:MB_RL1} which receive internal reward signals computed in the CX: comparing the agent's current compass heading $\kappa$ and PI home vector (with compass angle $Pi$) by a mechanism similar to the one described by \citet{lyuBuildingAllocentricTravelling2022}, a set of two reward signals $r_{l/r}$ is computed, encoding whether the nest is located to the left/right of the agents current heading direction.
These provide input for two sets of dopaminergic neurons $D_{l/r}$, assuming the role of external rewards in the canonical model
(Fig.~\ref{fig:MB_RL1}A).
However, they don't encode a rewarding experience coming from the environment, like the agent reaching a food source, but relate to an internal state of the agent, the home vector.
Since the latter is continually updated, the rewards are no longer sparse which makes learning considerably more efficient.
DANs convey copies of the respective reward signal to MBs in both hemispheres, giving rise to double opponent processes: In each hemisphere, $r_{r/l}$ serve as direct reinforcement to associate representations of the current view (encoded in KC activity) with populations of MBONs, $M_{l/r}$, corresponding to `steer left/steer right' responses respectively.
In either hemisphere, $M_{l/r}$ activity is integrated by downstream neurons, but with inverted signs, leading to competing `steer left' and `steer right' premotor commands from the left/right hemisphere, respectively\footnote{The implementation of the actual steering mechanism is again located in the CX, but we will not discuss this further here.
    For more detail refer to the original paper \citet{wystrachNeuronsPremotorAreas2023}}.
This double opponent architecture increases performance robustness as there are now two sets of MBONs that independently encode the appropriate behavioral response.

Evidently, the model implements an associative learning algorithm:
It does not involve computation of a TD error as proposed above, or even an RPE for the internal reward.
While it may be interesting to extend the model to use RPEs for reinforcement, a computation of TD errors would serve to achieve an erroneous objective: The formulation of the underlying MDP would imply that the agent's goal is to maximize the cumulative \emph{internal} reward.
Since the internal reward is higher (e.g. for steering right) the further off-target the agent is heading, maximizing it over time would lead to the opposite of the desired behavior.
It would be interesting to investigate if a model with inverted reward valences could be extended to an RL model for visual homing.
In the following, however, we will explore a different line of thought, sketching an MB/CX inspired RL model that integrates all three components of the insect navigation base model from Sec.
\ref{sec:spatial_rep_insect} - path integration, vector memories, and view memories - and links them to the behavioral objective of optimizing external rewards.
Finally, note that - in a liberal interpretation of RL terminology - we can classify the visual homing model as an \emph{off-policy} learning algorithm: During the learning phase, the agent's actions are assumed to follow an exploration strategy in agreement with observations \citep[using data from][]{jayatilakaChoreographyLearningWalks2018, wystrachVisualScanningBehaviours2014} of learning flights/walks performed by insects after emerging from their nest for the first time \citep[see][for a recent review]{collettInstinctLearningLearning2023}.

\begin{figure}
    \centering
    \includegraphics[width=0.5\textwidth]{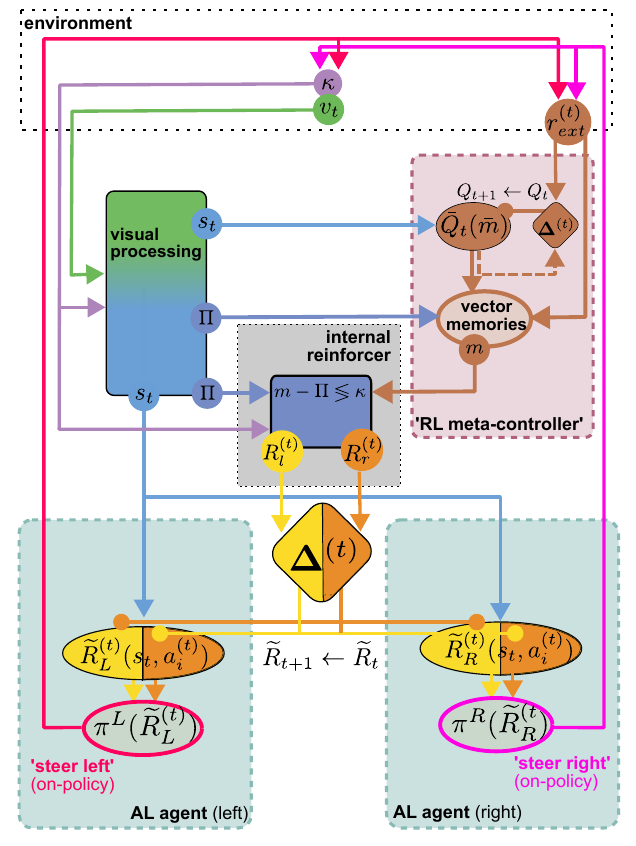}
    \caption{The visual homing model of Fig.~\ref{fig:MB_RL2} can conceptually be turned into a hierarchical MB/CX based RL full navigation model which roughly fits to the insect navigation base model.
        This can be done by replacing the off-policy controller in Fig.~\ref{fig:MB_RL2}B with an MB RL agent as an `RL meta-controller' (Fig.~\ref{fig:MB_RL1}D).
        As one additional component, it relies on a store of vector memories $m$, `snapshots' of the PI home vector $\Pi$, learned via direct associative reinforcement with an external reward at salient locations (see Sec.~\ref{sec:spatial_rep_insect} and Sec.~\ref{subsec:MBCXRL} for possible anatomical implementations) which now form the (discrete) action space of the TD-RL agent.
        A specific vector memory $\bar{m} \sim \pi(m|s)$ is selected as a navigational subgoal based on a policy learned from the same external reward as the vector memories on a state space of processed visual input $s_t$.
        The internal reward now encodes whether the agent is facing left/right with respect to the relative vector to the currently active vector memory selected by the meta-controller.
        This vector computation based on current $\Pi$, current heading $\kappa$, and active $m$ is assumed to be performed in the CX.
        The rest of the visual homing circuit is identical to Fig.~\ref{fig:MB_RL2}.
        Note that $s_t$ and $\Pi$ could conceptually be viewed as components of a joint representation of the visual input, with different components of the algorithm acting only on subspaces of this representation.
    }
    \label{fig:MB_RL3}
\end{figure}

\subsection{Towards an MB/CX based RL Model for (Insect) Navigation}
\label{subsec:MBCXRL}
The visual homing model discussed above will serve as a representative of models for view memory.
It obviates the need to decide when to store a view memory by learning continuous associations, as described above.
It makes a very explicit connection between view memories and the home vector, while other models only implicitly associate view memories with a motivational state \citep{webbInternalMapsInsects2019}.
However, taking into account the proposed mechanism for storing and recalling vector memories \citep{lemoelCentralComplexPotential2019}, one could broadly interpret the presently loaded vector memory as a motivational state (feeder vector loaded = `foraging motivation', no vector loaded = `homing motivation').
Loading the vector memory of a specific location effectively replaces the home vector in the visual homing model with a relative vector towards that location, theoretically allowing the agent to associatively learn visual homing relative to \emph{every location in the vector memory}, using internal rewards.
Vector memories on the other hand, are learned by one-shot associative learning from external reward, e.g. the presence of food.
As suggested by \citet{lemoelCentralComplexPotential2019}, this could be achieved by direct dopaminergic modulation of synapses between `vector-memory neurons' and the CPU4 integrator neurons in the CX.
Alternatively, specific populations of neurons, each of which conveys a specific vector in a phasor-like representation, might be activated.
If these hypothetical `vector-memory neurons' are, or receive excitatory input from, a subpopulation of MBONs, the pool of vector memories could in turn serve as the action space for an MB-based implementation of a TD-RL algorithm discussed above (Fig.~\ref{fig:MB_RL2}B).
This RL meta-controller learns a policy over sub-goals from the current view and \emph{external} reward, which may serve two purposes: Steering the agent towards the selected sub-goal using PI, and providing sub-goal-directed \emph{internal} reward for the low-level AL steering/homing controller.
Interestingly, however, due to the off-policy architecture of the AL controller, view association with respect to any (inactive) vector memory could be learned while homing towards another (active) one (using either views or PI), assuming a dedicated set of visual homing MBONs associated with each vector memory.
This would be consistent with our interpretation of vector memories as motivational states, which are often modeled by distinct MBON populations in the MB literature (see \citet{webbNeuralMechanismsInsect2016a}).
This would theoretically enable the agent to continually learn view associations with respect to all vectors stored in memory while using a specific one, or PI for steering.
If we further assume an anatomical link between `vector memory MBONs' and their corresponding set of `visual homing MBONs', the RL-meta controller could select previously visited vector locations as sub-goals for visual homing.
We argue that a unified MB/CX RL model along these lines would give rise to a remarkably versatile spatial representation for navigation.
Fig.~\ref{fig:MB_RL3} illustrates how the proposed model is built from the components discussed above.

\section{Discussion}
\label{sec:discussion}

In this paper, we presented reinforcement learning as a common framework to compare representations of space (Sec.
\ref{sec:latentspaces}) in current models for robot (Sec.~\ref{sec:spatial_rep_robo}) and insect navigation (Sec. \ref{sec:spatial_rep_insect}).
We argue that the insect navigation based model combines an \emph{explicit} vector map with a \emph{latent} spatial representation based on view memories.
This combination of explicit and latent spatial representations, evidently very successful for insects, may provide inspiration for future approaches for robot navigation.
We further proposed reinforcement learning as a novel framework for interpreting and expanding existing models for insect navigation.
In Sec.
\ref{sec:RL}, we analyzed how existing models of the insect MB circuit could implement temporal-difference (TD) RL algorithms, and proposed a circuit model with specific recurrent MBON$\rightarrow$DAN connections which would support TD-like reinforcement at the MBON$\rightarrow$KC synapse via differential temporal dynamics (Sec.~\ref{subsubsec:MB_RL})
In Sec.~\ref{sec:RL_insect_navigation}, we sketched how this could be used to build a full MB/CX inspired TD-RL navigation model, which has the potential to link the disjoint spatial representations of the insect navigation-based model.
To conclude, we will discuss aspects of such a hypothetical model with respect to spatial representations and RL-based robot navigation models more generally.

\subsection{Task Hierarchies, Intrinsic Rewards, and Exploration}
The model we outline here is, in a sense, a hierarchical RL (HRL) model: A metacontroller selects a sub-goal for a low-level controller, to provide the latter with dense internal rewards instead of operating on the sparse external reward landscape.
For similar reasons, hierarchical RL architectures also play an important role in robot navigation \citep{schmalstiegLearningHierarchicalInteractive2023,nachumDataEfficientHierarchicalReinforcement2018, haarnojaLatentSpacePolicies2018} to avoid `dimensional disaster'.
There are, however, crucial differences: In RL, the meta-controller usually provides \emph{intrinsic} rewards for the low-level agent, designed to facilitate the exploration of complex environments with sparse external rewards.
\citet{oudeyerHowCanWe2008} define a situation as  \emph{`intrinsically motivating [...] if its interest depends primarily on the collation or comparison of information from different stimuli and independently of their semantics, [...] understood in an information theoretic perspective, in which what is considered is the intrinsic mathematical structure of the values of stimuli, independently of their meaning'}, as opposed to extrinsic motivation.
Intrinsic rewards can be viewed as a formalization of curiosity, motivating the agent to explore unfamiliar terrain, and various approaches to model it exist in the RL literature \citep{,pathakCuriositydrivenExplorationSelfsupervised2017,burdaExplorationRandomNetwork2018,savinovEpisodicCuriosityReachability2019}.
In the proposed model, the internal reward \textendash{} facing left/right of a location related to external rewards \textendash{} is inherently \emph{extrinsic} and does not motivate the agent to explore, and the low-level controller is not an RL agent.
However, there are biologically plausible ways to include intrinsic reward in the model.
For example, \citet{sunDecentralisedNeuralModel2020} proposes a mechanism for switching between on-route and off-route navigation strategies based on visual novelty.
Such a signal could theoretically serve as an intrinsic reward in our model.
\subsection{A Unified Spatial Representation}
The proposed CX/MB inspired RL navigation model would fuse the disjoint spatial representations of the insect navigation base model \textendash{} an explicit vector map and latent directional cues (Fig.
\ref{fig:space_latent_rep}) \textendash{} into a unified latent graph-like representation which could be used in the absence of accurate PI information (while preserving the vector map for pure PI-based navigation).
The agent learns to select a high-level subgoal - a vector memory - based on the current view, from the high-level objective of external reward optimization using an MB/CX RL circuit.
This association defines a latent topological relation between the location of the current view and nodes represented by the vector memories: the learned policy will likely reflect distance since learning to visit distant goals would be less rewarding in the long run.
The low-level MB/CX AL homing circuit learns rough directional cues to steer the agent toward the selected goal.
Although both directional and distance information are therefore available to the agent, both are indirect and inaccurate, making the representation topological rather than geometric.

In light of the cognitive map debate, we can characterize an agent with such a spatial representation as \emph{`knowing where to go, on different spatial scales'}: It can infer from the current view both where it wants to go, i.e. which vector memory to load, and how to get there.
This would enable behavior commonly associated with a cognitive map.
For example, the agent could visually infer the direction of a novel shortcut from the present location A to a previously visited location B stored in the vector memory, since view memories in the vicinity of A have also been associated with left/right steering commands with respect to B.
By a change in the policy over vector memories, for example, because a previously closer food source C has been depleted, the agent could be incentivized to choose to steer along a novel route ('shortcut') towards B.

\subsection{Predictions, Memory Traces, and Internal World Models}
The previous example shows that our model could support surprisingly flexible navigation, adapting behavior locally depending on distant changes in the environment.
In RL, this kind of flexibility is traditionally associated with model-based algorithms: Unlike the model-free algorithms discussed above, they learn a model of the environment \textendash{} i.e. the transition Eq.~\eqref{eq:transition} and reward functions Eq.~\eqref{eq:reward} \textendash{} directly and then infer the optimal policy from them via Eq.~\eqref{eq:BellmanV1}.
Explicit knowledge about the environment allows them to plan out actions virtually in order to optimize the policy
(see \citealt{hafnerDeepHierarchicalPlanning2022} for a current application using Deep Hierarchical Planning).
This is particularly useful to adapt flexibly to changes in the environment, like changes in reward magnitudes (e.g. an empty feeder) or new navigational obstacles.
Model-based algorithms don't need to slowly and incrementally update their state evaluation by repeated exposure to the change in the environment, but can simply update their model of the environment and virtually plan a new route based on that update.
A neural implementation of such a \emph{predictive} model of the environment presupposes a state representation in the form of recurrent neural activity, like the (p)replay of spatial sequences observed in the mammalian hippocampus \citep{dragoiPreplayFuturePlace2011}.
The likely insect analog, spontaneous KC activity in the MB, would be challenging to access experimentally and has not yet been observed.
On the modelling side, \citet{weiLearningSparseReward2024} recently proposed the intriguing idea that adaptive axo-axonal gap junctions between KCs in the MB calyx could represent the state transitions of a learned transition function and thus support model-based RL computations in the MB.

On the other hand, a suitably predictive (latent) representation can produce similar flexible behavior.
E.g. \citet{russekPredictiveRepresentationsCan2017} show that the successor representation \citep{dayan1993improving} can link model-free TD methods to model-based behavior.
Abstract latent representations like in the SMT model (\citealt{fangSceneMemoryTransformer2019}, see Sec.~\ref{sec:spatial_rep_robo}) can also adapt flexibly to environmental changes without an explicit model: Adding an observation of a change in the environment to the scene memory would globally change the embedding of all other scene memories, and the latent representation could quickly adapt to reflect the new environment.
Another memory mechanism linking model-free TD-RL algorithms to seemingly \emph{predictive} behavior are \emph{eligibility traces} \citet{suttonLearningPredictMethods1988a}.
Instead of updating only values of the current (single timestep) state-action pair with the current reward, a trace of previous experiences is updated as well.
A current negative reward, e.g. related to an obstacle, would for example affect the agent's evaluation of an earlier state, causing it to adapt behavior early.
However, while the behavior looks predictive, it is in fact reactive: In order to learn to avoid a new obstacle, one would still have to experience the novel situation a couple of times.
\citet{wystrachRapidAversiveMemory2020} used a very similar concept of memory trace learning\footnote{This is essentially the AL analog to eligibility traces.}, mediated by KC activity traces in the MB, to show how desert ants learn to avoid new obstacles.

Our MB/CX RL model has the potential to enable fast adaptation to changes: As discussed in the previous paragraph, a small change in global high-level policy due to environmental changes (empty feeder) can lead to sudden change in the local, low-level behavior (steering towards a different goal):
The question would now be, how fast can the agent adapt the high-level policy to the environmental change?
Since local steering is handled by the low-level controller, and the agent learns a policy over vector memories on a relatively large spatial scale, the agent doesn't need to constantly reassess said policy, but can do so in larger intervals.
This would effectively increase the temporal scale of the MDP, and therefore reduce the number of intermediate states between rewarding experiences.
In this `sped-up' MDP with much denser rewards, policy adaptation due to a local change in the environment is propagated faster to distant states, through fewer intermediate states.
This effect could be further enhanced by a memory trace mechanism.
Taken together, this would enable the agent to quickly adapt local behavior based on distant changes in the environment, by updating its hierarchical, topological spatial representation of the world, without the need for a predictive internal world model.

\subsection{RL as a Normative Framework: Place-Cell like Activity in KCs?}
So far, we have treated the visual system as a static component of the model.
While this assumption is largely consistent with current knowledge about plasticity in the insect visual system, we can again expand the temporal horizon to evolutionary timescales and view the anatomical components of the insect MB, CX, and visual system in the light of end-to-end RL, as outlined in the discussion of Sec.~\ref{sec:spatial_rep_insect}.
Without going into anatomical details of the insect visual system here, a sensor-motor signal transduction pathway: retina $\rightarrow$ lamina $\rightarrow$ medulla $\rightarrow$ lobula / lobula plate $\rightarrow$ visual PNs $\rightarrow$ KCs $\rightarrow$ MBONs $\rightarrow$ (pre)motor neurons, could be modeled as a deep neural network implementing a DRL architecture, whose latent spatial representation would then correspond to KC activity (the input layer to the actual policy network, represented by the MB).
The learned representations could be indicative of actual spatial representations used by navigating insects, and provide guidance for experimental work.
For example, it would be interesting to see if such a model learns a grid-like spatial representation which would correspond to place-cell like KC activity.
This hypothesis is compatible with known sparse firing patterns of KCs, but otherwise speculative given current neurophysiological data.
It is also unclear how such a representation would fit into the existing model for insect navigation.
Since CX based PI is firmly established as a key component for insect navigation, it seems imperative to eventually include the CX in such an end-to-end RL model.
This would enable representations that embed PI-like components in a more complex latent space.
HOwever, current insect navigation models do not include CX$\rightarrow$KC connections and it is not straightforward how the CX would be integrated into a MB based end-to-end RL model.
It will also be challenging to strike a balance between expressive power of the network architecture \textendash{} essential for gaining new insights about possible representations \textendash{} and necessary constraints to match empirically known components of these representations, like the PI home vector.

\section*{Conflict of Interest Statement}

The authors declare that the research was conducted in the absence of any commercial or financial relationships that could be construed as a potential conflict of interest.

\section*{Author Contributions}

SL, DH, AV, and ADS: conceptualization; SL: writing - original draft; DH, AV, and ADS: writing - review \& editing; SL: visualization; AV and ADS: supervision \& funding acquisition.

\section*{Funding}
We gratefully acknowledge financial support from the BrainWorlds
Initiative at the University of Freiburg and the VolkswagenFoundation Momentum
Program (AZ 98692 to ADS).
We further acknowledge support by the Open Access Publication Fund of the University of Freiburg.
\section*{Acknowledgments}
We thank Joschka Boedecker, Paulina Friemann, Michael
J.
M.
Harrap and Christian Leibold for helpful
discussions.

\bibliographystyle{Frontiers-Harvard} %
\bibliography{LochnerReinforcementLearning}

\end{document}